\documentclass[twocolumn,showpacs,preprintnumbers,aps,amsmath,a4paper,amssymb,amsfont,superscriptaddress,xcolor=dvipsnames]{revtex4-1}
\pdfoutput=1

\usepackage[pdftex]{graphicx}
\usepackage{graphicx}
\usepackage{dcolumn}
\usepackage{bm}
\usepackage{float}
\usepackage[pdftex,colorlinks=true, citecolor = blue,urlcolor=black,linkcolor=black]{hyperref} 
\usepackage{microtype}
\usepackage{epstopdf} 
\usepackage{siunitx}
\usepackage{braket}
\usepackage{xcolor}
\usepackage{soul}
\usepackage{comment}

\DeclareSIUnit\micron{\micro\metre}
\DeclareSIUnit\mrad{\milli\rad}
\DeclareSIUnit\erec{\mathit{E}_{rec}}

\begin{document}
\title{ Floquet-driven crossover from density-assisted tunneling to enhanced pair tunneling}

\author{Nick Klemmer}
\thanks{These authors contributed equally.}
\author{Janek Fleper}
\thanks{These authors contributed equally.}
\author{Valentin Jonas}
\author{Ameneh Sheikhan}
\author{Corinna Kollath}
\author{Michael Köhl}
\author{Andrea Bergschneider}
\affiliation{Physikalisches Institut, University of Bonn, 53115 Bonn, Germany}

\begin{abstract}
We investigate the experimental control of pair tunneling in a double-well potential using Floquet engineering. We demonstrate a crossover from a regime with density-assisted tunneling to dominant pair tunneling by tuning the effective interactions. Furthermore, we show that the pair tunneling rate can be enhanced not only compared to the Floquet-reduced single-particle tunneling but even beyond the static superexchange rate, while keeping the effective interaction in a relevant range. This opens possibilities to realize models with explicit pair tunneling in ultracold atomic systems.
\end{abstract}

\maketitle
Understanding the complexity of correlated quantum matter from microscopic principles is a notoriously difficult challenge. One example of a bottom-up approach is the Hubbard model, which strives to explain macroscopic quantum states by the two elementary processes: single-particle tunneling $t$ between neighboring lattice sites and two-particle on-site interaction of strength $U$, see Figure \ref{fig:fig1}(a). Within this framework, one is able to explain the occurrence of Mott insulators \cite{Greif2016, Cheuk2016}, superconductors \cite{Scalapino1986} or quantum magnets \cite{Boll2016,Mazurenko2017,Bohrdt2021}, and their fundamental properties. However, the Hubbard model makes substantial simplifications in the description of matter: tunneling processes, for example, are always that of single particles and neither correlated
 tunneling of pairs nor density-assisted tunneling, where the tunneling depends on the occupation of the site, are explicitly accounted for. 
Instead, pair tunneling only arises implicitly as a second-order process at strong interactions ($|U|>t$). There, the interaction favors correlated  tunneling of single particles through a virtual state with a rate proportional to $\sim t^2/|U|$, see Figure \ref{fig:fig1}(c)-(e). This superexchange process as well as density-assisted tunneling have been observed with ultracold atoms in double-well potentials \cite{Fölling2007,Trotzky2008} and optical lattices \cite{Meinert2014,Jürgensen2014}.

It has been theoretically predicted that explicit pair tunneling as well as density-assisted tunneling of a significant strength give rise to novel quantum phases \cite{Robaszkiewicz1999,Japaridze2001,Kraus2013,Lang2015,Lisandrini2022} beyond the standard Hubbard model. 
An effective, density-assisted tunneling process is required for mapping a three-orbital model to a single-orbital model \cite{Jiang2023}, and explicit pair tunneling as realized in the Penson-Kolb-Hubbard model \cite{Penson1986}, can induce, for instance, $\eta$-pairing phases \cite{Robaszkiewicz1999,Japaridze2001}. 
Moreover, in the absence of single-particle tunneling an explicit pair tunneling has been predicted to lead to topologically non-trivial phases by the induced symmetries \cite{Kraus2013,Lang2015,Lisandrini2022}. Such phases are yet elusive in static realizations since pair tunneling is usually suppressed significantly as compared to single-particle tunneling by the larger mass of the pairs. 

In recent years, Floquet engineering \cite{Goldman2014, Bukov2015, Eckardt2017} has been developed to control parameters and quantum phases in ultracold atom systems \cite{Lignier2007, Meinert2016, Görg2019, Schweizer2019, Weitenberg2021}. The key idea of Floquet engineering is to periodically drive a system parameter, such as the potential bias between neighbouring lattice sites, at a high frequency. In an effective, time-independent description this leads to a renormalized tunneling $t_\textrm{eff}$ and interaction strength $U_\textrm{eff}$, see Figure \ref{fig:fig1}(b). Using Floquet engineering, researchers have demonstrated to suppress single-particle tunneling \cite{Lignier2007}, induce density-assisted tunneling \cite{Meinert2016,Görg2018}, or invert the sign of magnetic correlations \cite{Görg2018}, to name just a few examples, and it seems worthwhile to explore its potential benefits with regards to the control of pair tunneling, too. However, the Floquet engineering comes at a cost: for a resonant Floquet drive the effective interaction energy is nulled \cite{Messer2018}. 

 \begin{figure}
        \centering
        \includegraphics[width=\columnwidth]{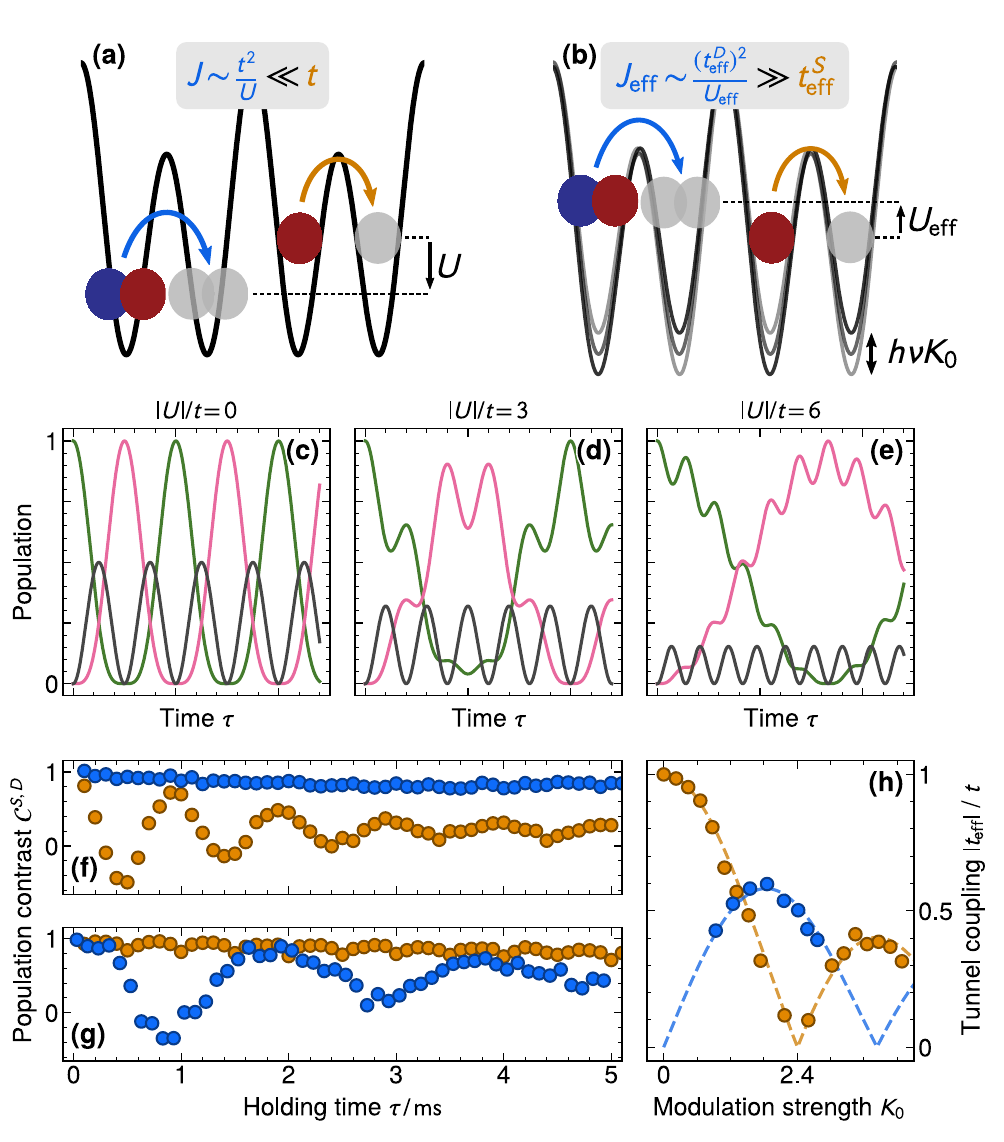}
        \caption{Controlling tunneling dynamics in static and driven double wells. (a) In the static double well, single-particle tunneling $t$ and the on-site interaction $U$ lead to the two-particle tunneling $J \ll t$. (b) In a driven double well with driving frequency $\nu$ and amplitude $h\nu K_0$, the effective pair-tunneling rate can exceed the effective single-particle  tunneling rates $J_\text{eff}\gg t^S_\text{eff}$.
        (c)-(e) Tunneling dynamics with negligible to strong interaction. In a non-interacting system at half filling (c), the pair tunnels from the left lattice site (green) to the right lattice site (pink) by going through the split state (black). For larger interactions (d,e), the population of the split state is significantly suppressed.       
        Measured tunneling dynamics at $U/t\approx-9$ for the static  (f) and resonantly driven double well (g) with driving frequency $\nu=\left|U\right|/h$ and amplitude $K_0=2.4$. The population contrast between left and right lattice sites is measured for quarter-filled $\mathcal{C}^S$ (orange) and half-filled double wells $\mathcal{C}^D$ (blue). (h) Effective tunnel coupling of resonantly driven double wells extracted from tunneling dynamics. In quarter-filled (half-filled) double wells, the tunneling follows a 0th-order Bessel function $\mathcal J_0(K_0)$ (1st-order Bessel function $\mathcal J_1(K_0)$) in orange (blue), respectively.
        }
        \label{fig:fig1}
\end{figure}
In this work, we demonstrate experimentally a crossover from a density-assisted tunneling to an enhanced pair-tunneling regime via Floquet engineering by tuning the driving frequency with respect to the static on-site interaction. First, we study density-assisted tunneling in effectively non-interacting driven double wells while fully suppressing single-particle tunneling. Then, we introduce effective interactions and demonstrate the enhancement of pair tunneling by suppressing the population of the intermediate state in which the populations are split between the sites (cf. Figure~\ref{fig:fig1}(c)). An especially intriguing finding is that for large effective interactions the pair tunneling rate can be enhanced significantly above the superexchange rate of the corresponding time-independent effective description of the double well.

We perform our experiment using a quantum degenerate sample of $^{40}$K atoms in the lowest two hyperfine states, denoted as $\ket{\uparrow}$ and $\ket{\downarrow}$. The atoms are confined in a three-dimensional optical lattice formed by off-resonant laser beams. Along the $x$-direction, the optical lattice is bichromatic  and composed of two retro-reflected laser beams of an infrared (\SI{1064}{\nano\meter}) and a green (\SI{532}{\nano\meter}) wavelength in order to form a chain of tunable double wells. We achieve tunability of the barrier heights within and between the double wells as well as of the potential bias between the sites of each double well by controlling both laser powers and frequencies \cite{Fölling2007,Pertot2013}. In all experiments discussed here,  
we focus on isolated double wells by suppressing the tunneling amplitude between double wells using a depth of \SI{15}{\erec} (\SI{12}{\erec}) for the \SI{1064}{\nano\meter} (\SI{532}{\nano\meter}) lattice while freezing the dynamics along $y$- and $z$-directions with lattice depths of \SI{60}{\erec} and \SI{110}{\erec}, respectively. Here, $E_\mathrm{rec}=h/(2m\lambda^2)$ denotes the recoil energy for atoms with mass $m$ in an optical lattice of wavelength $\lambda$ and $h$ is Planck's constant.
In this work, we choose $t/h = \SI{480}{\hertz}$ and an interaction strength of $U/t\approx-9$, if not stated otherwise.

We begin our experimental sequence by loading a balanced spin mixture into the lowest Bloch band of the infrared lattice along the $x$--direction. This results in an occupation of at most one $\ket{\uparrow}$ and one $\ket{\downarrow}$ atom per lattice site. Subsequently, the green lattice is adiabatically ramped up to \SI{25}{\erec} at a significant potential bias $|\Delta_0| \gg t,\left|U\right|$ between the left and right site of the resulting double-well array to ensure population of only the left sublattice site. This produces either half-filled double wells (a pair with one $\ket{\uparrow}$ and one $\ket{\downarrow}$ atom), quarter-filled double wells (with either a $\ket{\uparrow}$ or a $\ket{\downarrow}$ atom) or empty double wells in rare cases. In the first experiment, we then induce tunneling oscillations between left and right sites by removing the potential bias and simultaneously unfreezing the lattice within \SI{100
}{\micro\second} in a diabatic fashion. 
During the following time evolution, the system can be described by an array of double-well Hamiltonians
\begin{align*}
        \hat H=&-t \sum_{\sigma}\left(\hat{c}^\dagger_{L\sigma}\hat{c}^{}_{R\sigma}+h.c.\right)\\ &+U\sum_i\hat{n}_{i\uparrow}\hat{n}_{i\downarrow} +\sum_{\sigma}\frac{\Delta_0}{2}(\hat{n}_{L\sigma}-\hat{n}_{R\sigma}),
\end{align*}
with the fermionic raising (lowering) operators $\hat{c}_{i\sigma}$ ($\hat{c}^\dagger_{i\sigma}$), and the number operator $\hat{n}_{i\sigma}=\hat{c}^\dagger_{i\sigma}\hat{c}_{i\sigma}$ at sublattice site $i=L,R$ in spin state $\sigma =\, \uparrow,\downarrow$. 
After a holding time $\tau$, we freeze the dynamics and detect the system by resolving single and double occupancies at left and right sites of the double wells.
To resolve the site occupation, we transfer atoms at left (right) sites to the first (third and fourth) Bloch band of the superlattice and then employ an adiabatic band-mapping technique \cite{Sebby-Strabley2007,Fölling2007,Pertot2013}. From the measured Brillouin-zone occupations $N_1$  and $N_{34}=N_{3}+N_4$, we extract the site occupation-resolved population contrast $\mathcal{C}^{S}$ ($\mathcal{C}^{D}$) between left and right lattice sites with single occupancy (double occupancy), where $\mathcal C =\left(N_{1}-N_{34}\right)/\left(N_{1}+N_{34}\right)$.

The tunneling dynamics in the static double well with strong attractive interaction is shown in Figure~\ref{fig:fig1}(f). While for quarter-filled double wells, the atoms tunnel between the lattice sites, an almost constant population contrast is measured for the half-filled double wells due to the small pair-tunneling rates. The observed dephasing of the oscillation mainly stems from the Gaussian envelope of our lattice laser beams, resulting in approximately \SI{20}{\percent} variation in the tunneling strength $t$ and \SI{10}{\percent} in the interaction strength $U$, as well as a small, position-dependent bias $\Delta_0$ between adjacent lattice wells.

A periodic modulation of the system can fundamentally alter the tunneling dynamics \cite{Lignier2007,Eckardt2009,Schweizer2019,Weitenberg2021}.  
In particular, a high-frequency drive close to resonance with the interaction energy $\ell\nu \approx \left|U\right|/h \gg t/h$, where $\ell$ is an integer and denotes the harmonic order, alters the dynamics in quarter- and half-filled double wells in different ways.
For the half-filled double wells, the near-resonant drive modifies the interaction energy $U_\mathrm{eff} = U+\ell h\nu$, for $U<0$, and introduces density-assisted tunneling with amplitude $t^{(\ell)}_\mathrm{eff}= t \mathcal{J}_{\ell}(K_0)$ \cite{Liberto2014,Meinert2016,Messer2018}. At the same time, in quarter-filled double wells, the modulation frequency is the highest energy scale and reduces the single-particle tunneling amplitude to $t^S_\mathrm{eff} = t \mathcal{J}_{0}(K_0)$ \cite{Eckardt2009,Keay1995}.
We will show that this interplay of effective interactions and density-assisted tunneling can introduce significant pair tunneling $\hat{c}^\dagger_{R\uparrow}\hat{c}^\dagger_{R\downarrow}\hat{c}_{L\uparrow}\hat{c}_{L\downarrow}$ with amplitude $J_\mathrm{eff}$ independent of $t^S_\mathrm{eff}$.

We realize the periodically modulated system by starting with the preparation as in the static case discussed before where particles are localized at the left sublattice sites. Then, we adiabatically ramp on a periodic modulation of the double-well bias $\Delta_0\rightarrow \Delta(\tau)=\Delta_0+h\nu K_0\cos(2\pi\nu\tau)$ at a frequency $\nu$ and amplitude $h\nu K_0$. Subsequently, we unfreeze and measure the tunneling dynamics of the Floquet-driven double well \cite{Grifioni1998,Kierig2008,Chen2011,Goldman2015} by extracting $\mathcal C^{S,D}(\tau)$. The resulting dynamics for a resonant drive $\nu=\left|U\right|/h\gg t$ and $K_0=2.4$ is shown in Figure \ref{fig:fig1}\,(g). In contrast to the static case, the quarter-filled double wells dynamically localize while for the half-filled double wells, the resonant drive balances the interaction energy and introduces density-assisted tunneling with about half the frequency as for the static system. In Figure~\ref{fig:fig1}(h), we show the different effective tunneling amplitudes of quarter- and half-filled double wells for different driving strength following the Bessel functions of different orders. This shows that the periodic drive can be employed in order to tune the ratio of the single-particle to density-assisted tunneling over a wide range.

In the following, we study the crossover from density-assisted tunneling to pair-tunneling by introducing effective interactions. To this end, we detune the driving frequency from resonance while suppressing the single-particle tunneling via the driving amplitude $K_0=2.4$. The impact of effective interactions and higher order tunneling processes on the dynamic signals is shown in Figure~\ref{fig:fig2}. For an effectively non-interacting system with $\nu\approx \left|U\right|/h$ (see Figure~\ref{fig:fig2}(a)) the oscillation dynamics exhibit a single characteristic frequency. Upon introducing a small detuning of the driving frequency from the resonance, a beating becomes apparent in the contrast measurement (see Figure~\ref{fig:fig2}(b)-(e)). This beating becomes more pronounced with increasing detuning and is accompanied by a decrease in oscillation amplitude. When the driving frequency is approximately $\nu\approx U/(2h)$ (cf. Figure~\ref{fig:fig2}(e)), the oscillation amplitude increases again and the characteristic signature of frequency beating is less pronounced. This indicates that another effectively non-interacting system is realized by the second-harmonic resonance condition \cite{Schweizer2019,Görg2019}.

\begin{figure}
        \centering
        \includegraphics[width=\columnwidth]{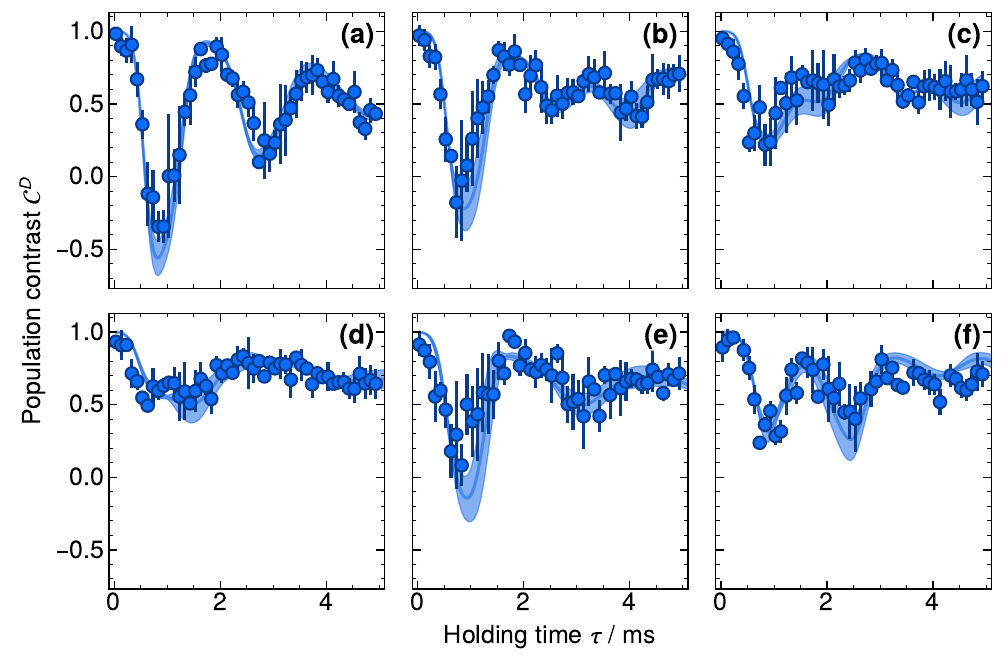}
        \caption{Measured dynamics in driven, half-filled double wells and attractive interaction $U/h\approx \SI{-4}{\kilo \hertz}$ ($U/t \approx -9$). The measured population contrast $\mathcal{C^D}$ is shown for driving amplitude $K_0  =  2.4$ and driving frequencies: \SIlist{4.05; 3.9; 3.6; 2.7; 2.25; 2.05}{\kilo \hertz} (a to f). The fitted time evolution of the global Hamiltonian is shown as solid blue line and the shaded regions indicate the $1\sigma$ confidence intervals.}
        \label{fig:fig2}
\end{figure}

We compare our experimental data to theoretical calculations 
using Floquet theory to determine the time-independent effective Floquet Hamiltonian for a near-resonant drive \cite{Goldman2014,Bukov2015,Mikami2016}. For this purpose, we employ a high-frequency expansion of the Floquet Hamiltonian up to the third order in $1/\nu$, while incorporating the Bessel functions $\mathcal{J}_{j}(K_0)$ for $j=0,1,2,3$ \cite{Mikami2016,Desbuquois2017}.
In order to account for the inhomogeneous intensity distribution of our optical lattice in our theoretical analysis, we calculate the spatially dependent Floquet and Hubbard parameters $K_0(\boldsymbol{x})$, $t(\boldsymbol{x})$, $U(\boldsymbol{x})$, and $\Delta_0(\boldsymbol{x})$. To obtain $t$, we compute the Wannier functions using the band-projected position operator method \cite{Kivelson1982,Uehlinger2013} and use this to calculate $U$ based on the $s$--wave scattering length $a$ \cite{Schneider2009,Chen2020}.
With these parameters, we solve the different Floquet-driven double-well Hamiltonians and 
compute the time- and position-dependent contrast $\mathcal{C}^D(\boldsymbol{x},\tau)$. We then perform 
a density-weighted average over the inhomogeneous density distribution resulting in the global observable $\mathcal{C}^D(\tau)$. To account for experimental fluctuations, this contrast is then fitted to our experimental data leaving the scattering length $a$ and the green lattice potential depth at the trap center as  fit parameters. 
The fitted results are presented as solid lines in Figure~\ref{fig:fig2}, with the $1\sigma$ confidence interval due to residual uncertainties of the Floquet and Hubbard parameters indicated by a shaded region. The numerical data accurately replicates the dynamic features observed in our experimental data as well as a pronounced dephasing. The fitted scattering lengths and lattice depths are in good agreement with our experimental uncertainties.

\begin{figure}
        \centering
        \includegraphics[width=\columnwidth]{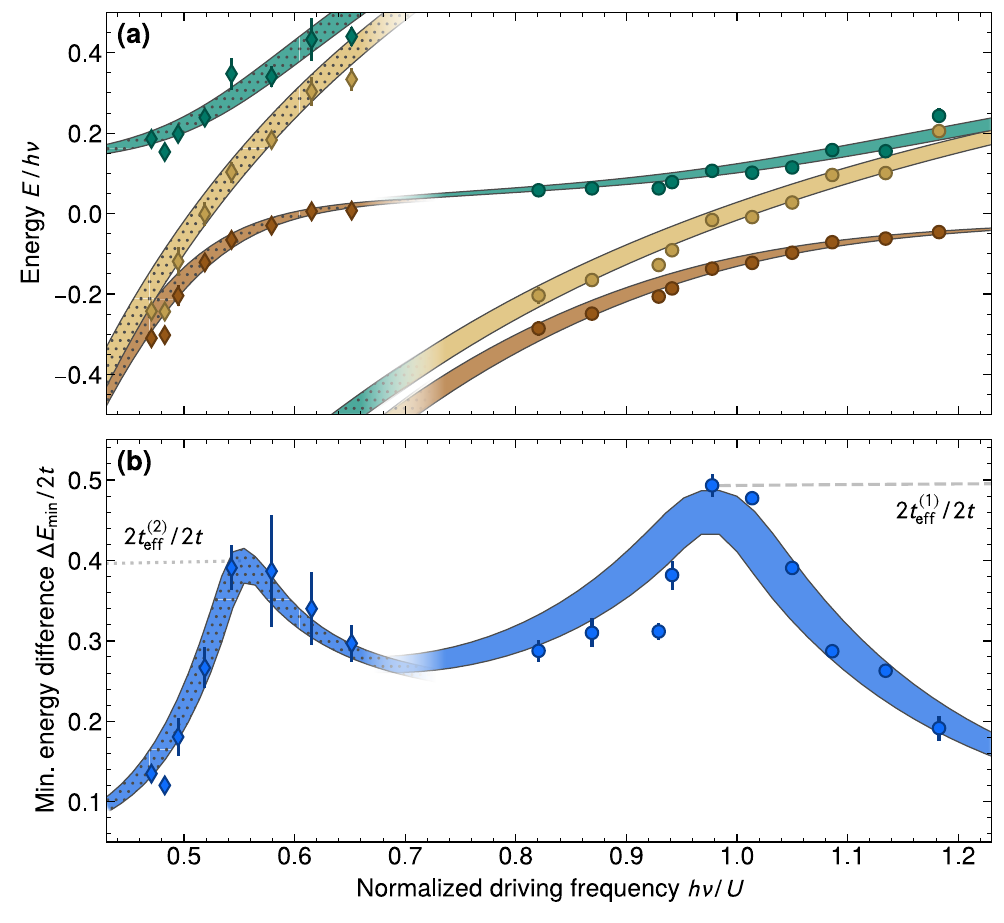}
        \caption{Spectrum of the effective 
        Hamiltonian for half-filled double wells. (a) Extracted energy eigenvalues $E/(h\nu)$ of the fitted Hamiltonian at the trap center for various driving frequencies $h\nu /U$ at a driving amplitude $K_0=2.4$ for the first (second) harmonic near-resonant case as circles (diamonds).  (b) Minimal energy difference $\Delta E_\mathrm{min}/(2t)$ between two energy eigenstates of the spectrum compared to the single-particle tunneling (orange line). The density-assisted tunneling amplitudes for the effectively non-interacting systems $t^\ell_\mathrm{eff}$ are indicated as dashed lines. The $1\sigma$ confidence interval of the first (second) harmonic near-resonant theory is shown as a shaded (hatched) region applying to both subplots.
        }
        \label{fig:fig3}
\end{figure}

To gain valuable insight on the effective processes that cause the observed dynamics, we extract the energy spectrum of the 
fitted effective Hamiltonian at the trap center for half-filling. This procedure provides three relevant energy eigenvalues per driving frequency, shown in Figure~\ref{fig:fig3}(a), and a fourth eigenvalue, which is constant zero due to a lack of coupling and hence omitted from display. 
The rescaled energy spectrum $E/h\nu$ as a function of the normalized driving frequency $h\nu/\left|U\right|$ reveals two distinct avoided crossings at $\nu \approx \left|U\right|/h$ and $\nu \approx \left|U\right|/(2h)$ corresponding to the two lowest harmonics. 
At the avoided crossings the system is effectively non-interacting and the energy gap between the eigenstates corresponds to the induced density-assisted tunneling $2\,t^{(\ell)}_\mathrm{eff}$.
Above resonance, the emerging effective repulsive interaction strength $U_\mathrm{eff}\approx U+\ell h\nu>0$ pushes the upper two energy branches up in an approximately linear fashion due to their increasing contribution of double occupancies. Similarly, below resonance, the energy of the lower two branches is decreased by the attractive interaction. 

We further analyze the effective processes by extracting the minimal energy difference $\Delta E_\mathrm{min}/(2t)$ between two eigenstates in the spectrum (see Figure \ref{fig:fig3}(b)).
For effectively non-interacting systems at resonance, this quantity corresponds to the 
density-assisted tunneling amplitude $2\,t^{(\ell)}_\mathrm{eff}$ that follows to lowest order the Bessel functions  $\mathcal{J}_{\ell}(K_0)$ depending on the harmonic order $ \ell$ as indicated by the gray dashed lines. Far from resonance, the arising large effective interactions suppress the breaking of the pairs and the minimal energy gap corresponds to the effective pair tunneling $|J_\mathrm{eff}|$ which approaches the superexchange rate $4\,t_\text{eff}^2/U_\text{eff}$ for large interactions. Note, that for the chosen parameters both tunneling processes are larger than the effective single-particle tunneling (orange line in Figure \ref{fig:fig3}(b)), which enables the realization of models with dominant higher-order tunneling processes. 

\begin{figure}
        \centering
        \includegraphics[width=\columnwidth]{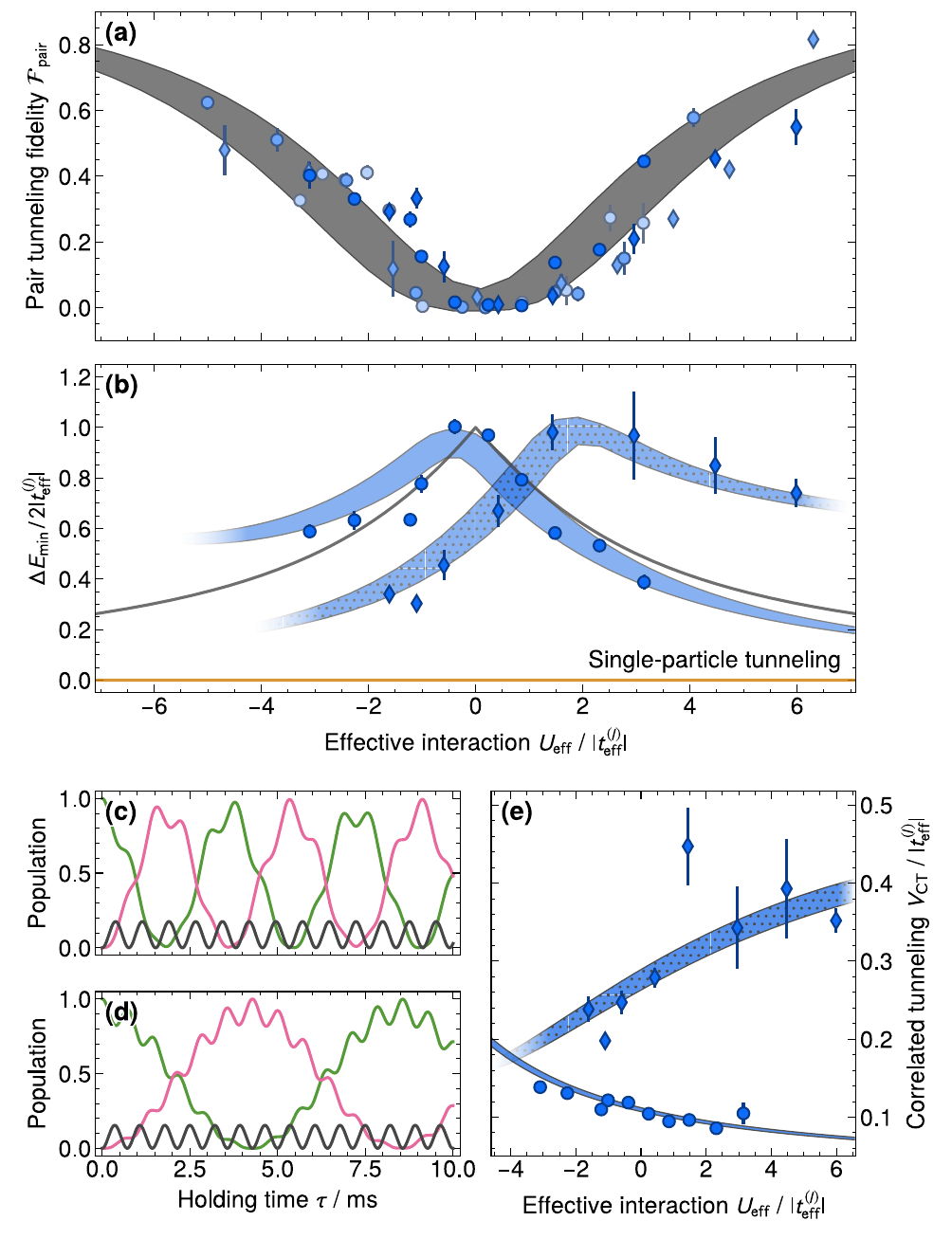}
        \caption{(a) The pair-tunneling fidelity $ \mathcal{F}_\mathrm{pair}$ is determined as a function of the effective interaction strength $U_\mathrm{eff}/|t_\mathrm{eff}^{(\ell)}|$ for various absolute interaction strengths $U/t\approx-9$ (dark blue), $U/t\approx-18$ (blue) and $U/t\approx-6$ (light blue) as circles (diamonds) for the first (second) harmonic resonance. (b) In the driven double well with harmonic order $\ell=2$ and effective interaction strength $U_\mathrm{eff}/|t^{(2)}_\mathrm{eff}|=6$ the pair tunnels from the left lattice site (green) to the right lattice site (pink) with only a reduced population of the split states (black) and higher frequency compared to the static double well (c).  (d) The extracted correlated tunneling rate $V_\mathrm{CT}/|t_\mathrm{eff}^{(\ell)}|$ is given as a function of the effective interaction strength  $U_\mathrm{eff}/t_\mathrm{eff}^{(\ell)}$ for $U/t=-9$, where the circles (diamonds) show the harmonic order $\ell=1$ ($\ell=2$). For all subplots, the $1\sigma$ confidence interval of the first (second) harmonic near-resonant theory is shown as a shaded (hatched) region.} 
        \label{fig:fig4}
\end{figure}

To map out the crossover from density-assisted tunneling to the pair tunneling, we quantify the pair splitting by the time-averaged population of split pairs $A_\mathrm{split} = \overline{\left|\left\langle LR| \Psi(\tau)\right\rangle\right|^2+\left|\left\langle RL| \Psi(\tau)\right\rangle\right|^2}$. From this, we extract the pair-tunneling fidelity $ \mathcal{F}_\mathrm{pair}=1-4 A_\mathrm{split}$ of the driven system (see Figure~\ref{fig:fig4}(a)). 
As expected, the pair tunneling fidelity increases with the effective interaction strength.
For ($\left| U_\mathrm{eff}/ t_\mathrm{eff}\right| > 5$), this fidelity exceeds \SI{60}{\percent}, rendering pair tunneling the dominant process in the time evolution of the system.

The pair-tunneling rate in the driven system can be significantly enhanced compared to its static counterpart where the rate is given by the superexchange. This is caused by a driving-induced enhancement of the correlated tunneling amplitude $V_\mathrm{ct}$ in the effective Hamiltonian, which describes the explicit hopping of pairs. $V_\mathrm{ct}$ contributes to the pair tunneling in addition to the superexchange with  $J_\mathrm{eff} \approx 4 \, t_\mathrm{eff}^2/U_\mathrm{eff}+ \left(-1\right)^\ell 2\,V_\mathrm{ct}$ for large effective interactions (see Supplemental Material \cite{SOM}). The harmonic order determines whether the pair tunneling is enhanced or suppressed by the correlated tunneling, which becomes apparent by the asymmetry of the curves in Figure\,\ref{fig:fig3}(b).

We extract the matrix element for the correlated pair tunneling $V_\mathrm{ct}$ from the fitted effective Hamiltonian (see Figure\,\ref{fig:fig4}(d)). The values in the range of up to 50\% of the effective density-assisted tunneling represent an enhancement by two orders of magnitude compared to the static situation and agree with the theoretical expectations (shaded regions in Figure\,\ref{fig:fig4}(d)). In the high-frequency expansion \cite{SOM}, $V_\mathrm{ct}$ scales in first order inversely with the drive frequency  resulting in an increase for lower $\nu$. This is the main contribution for $V_\mathrm{ct}$ at $\ell=1$, which increases for lower effective interactions. For the second harmonic order $\ell=2$, the lower drive frequency $\nu \approx U/(2h)$ leads to a larger $V_\mathrm{ct}$ and the higher-order terms enhance it at effective repulsive interactions. In contrast to the superexchange, larger effective interactions do not lead to a decrease of $V_\mathrm{ct}$, rendering correlated pair tunneling and the superexchange on par for $\ell=2$ at the largest repulsive effective interaction.
We illustrate this effect of the driving-enhanced pair tunneling in Figures \,\ref{fig:fig4}(b) and \ref{fig:fig4}(c) with the simulation of the pair-tunneling dynamics including the correlated tunneling term and for superexchange only, respectively. This demonstrates the increase of the tunneling rate by the periodic drive.

In conclusion, we have utilized Floquet engineering to realize a crossover from density-assisted tunneling to pair tunneling in an effectively interacting system, while simultaneously suppressing single-particle tunneling. We could further demonstrate that the realized pair tunneling rate is significantly enhanced beyond the expectations from superexchange processes in their static equivalent. This allows for the realization of models with dominating higher-order tunneling processes \cite{Penson1986} together with suppressed or negligible single-particle tunneling, which is not possible in the static Hubbard model.

C.K. and A.S. would like to thank Simon Bauer and Friedrich Hübner for their early contributions to this work. This work has been supported by Cluster of Excellence Matter and Light for Quantum Computing (ML4Q) EXC 2004/1–390534769 and the Deutsche Forschungsgemeinschaft (Project No.\,277625399 - TRR 185 (B3, B4), Project No.\,277146847 - CRC 1238 (C05), and   INST 217/1013-1 FUGG). This research was also supported in part by grant NSF PHY-1748958 to the Kavli Institute for Theoretical Physics (KITP).

\emph{Data availability} The supporting data for this article are openly available at Zenodo \cite{Zenodo}.”


%

\clearpage
\section*{Supplementary Material}

\subsection{Higher-band corrections to the static double-well Hamiltonian}
We study two interacting fermions of opposite spin in a double-well potential in a tight-binding picture \cite{Desbuquois2017}. For this purpose, 
we introduce the field operator \cite{Tarruel2018} 
\begin{equation}
    \hat{\psi}_\sigma(\boldsymbol{x}) = \sum_{i=\{L,R\}} w_i(\boldsymbol{x})\,\hat{c}_{i\sigma},
\end{equation}
with the fermionic annihilation operator $\hat{c}_{i\sigma}$ and the single-band Wannier functions $w_i(\boldsymbol{x})$ that we calculate as eigenfunctions of the band-projected position operator.
From this operator, we calculate the double-well Hamiltonian in the Fock-basis $\left\{ \ket{\uparrow\downarrow,0},\ket{\uparrow,\downarrow},\ket{\downarrow,\uparrow},\ket{0,\uparrow\downarrow}\right\}$ as
\begin{equation}\label{sup:eq1}
    H_\mathrm{dw} = 
    \begin{pmatrix}
        U & -t & t & 0 \\
        -t&0 & 0 &-t \\
        t &0 &0&t \\
        0&-t & t & U \\
    \end{pmatrix} + H_\mathrm{corr},
\end{equation}
with the single-particle tunneling $t$ and the on-site interaction $U$.
Here, the single-particle tunneling is calculated from the non-interacting Wannier functions
\begin{equation}
      t = - \int d\boldsymbol{x}\, w_L(\boldsymbol{x})\left[-\hbar^2\nabla^2/2m+V(r)\right]w_R(\boldsymbol{x})+\delta t,
\end{equation}
with the interaction-dependent correction \cite{Jürgensen2014} 
\begin{equation}
    \delta t = \frac{4\pi\hbar^2a}{m} \int d\boldsymbol{x}\, |w(\boldsymbol{x})|^2 \cdot w_L(\boldsymbol{x})w_R(\boldsymbol{x}),
\end{equation}
and the scattering length $a$.
Note that for the chosen lattice parameters and interaction strengths this correction is small $\delta t / t= \SI{4\pm1e-2}{}$.
To determine the on-site interaction $U$, we calculate the interaction energy of a corresponding anisotropic harmonic oscillator \cite{Chen2020} 
with a correction that accounts for the lattice anharmonicities \cite{Schneider2009} 
\begin{equation}
    U = U_\mathrm{HO}\cdot \frac{\int d\boldsymbol{x}|w(\boldsymbol{x})|^4}{\int d\boldsymbol{x}|\phi_G(\boldsymbol{x})|^4},
\end{equation}
with the eigenfunctions of the harmonic oscillator $\phi_G$.

Considering higher bands of the underlying optical superlattice gives the correction to the double-well Hamiltonian
\begin{equation}
    H_\mathrm{corr} = 
    \begin{pmatrix}
        0 & 0 & 0 & V_\mathrm{ct} \\
        0 &V_\mathrm{nn} &-V_\mathrm{de} &0 \\
        0 &-V_\mathrm{de}&V_\mathrm{nn}&0 \\
        V_\mathrm{ct}&0 & 0 & 0 \\
    \end{pmatrix},
\end{equation}
with the nearest-neighbor interaction $V_\mathrm{nn}$, the direct-exchange interaction $V_\mathrm{de}$, and the correlated tunneling $V_\mathrm{ct}$.
The latter describes a coherent hopping of pairs with amplitude 
\begin{equation}
    V_\mathrm{ct} = \frac{4\pi\hbar^2a}{m} \int d\boldsymbol{x} |w_L(\boldsymbol{x})|^2\cdot |w_R(\boldsymbol{x})|^2.
\end{equation}
However, the amplitude of the correlated tunneling in the realized static double wells is negligible $V_\mathrm{ct}/t\leq\SI{8.4\pm0.3e-4}{}$.
Note that the amplitudes of the nearest-neighbour interaction and direct-exchange interaction are identical to the correlated tunneling amplitude $V_\mathrm{nn}=V_\mathrm{de}=V_\mathrm{ct}$ and are therefore also neglected during the study of static double wells $H_\mathrm{corr}\sim0$.

The eigenstates of this system are conveniently discussed in a new basis that consists of a singlet state $\ket{s}$, a triplet state $\ket{t}$ and two states of double-occupancies with different parity $\ket{d_{\pm}}$, that are defined as
\begin{equation}
    \begin{aligned}
        \ket{s} & = \frac{1}{\sqrt{2}} (\ket{\uparrow,\downarrow}-\ket{\downarrow,\uparrow})\\
        \ket{t} & = \frac{1}{\sqrt{2}} (\ket{\uparrow,\downarrow}+\ket{\downarrow,\uparrow})\\
        \ket{d_+} & = \frac{1}{\sqrt{2}} (\ket{\uparrow\downarrow,0}+\ket{0,\uparrow\downarrow})\\
        \ket{d_-} & = \frac{1}{\sqrt{2}} (\ket{\uparrow\downarrow,0}-\ket{0,\uparrow\downarrow}).\\
    \end{aligned}
    .
\end{equation} 
The triplet state and the negative parity doublon state are eigenstates of the balanced double well with energies $E_t=0$ and $E_{d_-}=U$. The other two eigenstates are the ground state and the most-excited state which are superpositions of the singlet state and the even parity doublon state with energies $E_{g,e}=U/2\mp\sqrt{4t^2+U^2/4}$.
Note that the triplet state does not couple to the other states of the double well and is therefore neglected in the discussion of this work. 

Of particular interest in this work is the pair tunneling rate $J$, defined as the energy difference between the two eigenstates that consist dominantly of pairs. These states are $\ket{d_-}$ and the ground (excited) state for attractive (repulsive) interactions resulting in 
\begin{equation}
    J = E_{g,e}-E_{d_-}=  - U/2\pm\sqrt{4t^2+U^2/4}.
\end{equation}
For strong interactions $|U|\gg t$, this pair tunneling approaches the superexchange $|J|\sim 4t^2/U$.

\subsection{Near-resonantly driven double wells}
We study interacting double wells with a periodic modulation of the on-site energy.
This driven system can be described by the time-dependent Hamiltonian
\begin{equation}\label{sup:eq2}
    H(\tau) = H_{dw}+\hat{V}(\tau),
\end{equation}
with the interacting double-well Hamiltonian of eq.\,(\ref{sup:eq1}) and the modulated on-site energy
\begin{equation}
    \hat{V}(\tau)=h\nu K_0\cos(2\pi\nu\tau)\,h_\Delta,\;\;h_\Delta=
    \begin{pmatrix}
        1&0&0&0\\
        0&0&0&0\\
        0&0&0&0\\
        0&0&0&-1
    \end{pmatrix},
\end{equation}
with the driving frequency $\nu$ and dimensionless driving amplitude $K_0$.

We are interested in the time-independent effective Hamiltonian for a near-resonant drive $\ell h\nu\approx U\gg t$ of harmonic order $\ell$ \cite{Bukov2015, Desbuquois2017}. 
First, we transform eq.\,(\ref{sup:eq2}) to the co-rotating frame with $U/h$ and $\nu$ resulting in 
\begin{equation}\label{cpt5:21}
    H_\mathrm{rot}(\tau) = 
    \begin{pmatrix}
    U-\ell h \nu        & -t_+(\tau,\ell)      & t_+(\tau,\ell)      & 0               \\
    -t^\ast_+(\tau,\ell)& 0                    & 0                   & -t_-(\tau,\ell) \\
    t^\ast_+(\tau,\ell) & 0                    & 0                   &t_-(\tau,\ell)   \\
    0                   & -t_-^\ast(\tau,\ell) & t^\ast_-(\tau,\ell) & U - \ell h \nu
    \end{pmatrix}
\end{equation}
with $t_\pm(\tau,\ell) = t \exp\left[i(\pm2\pi\ell \nu \tau + K_0\sin(2\pi\nu\tau))\right]$.
Then, we compute the effective Floquet Hamiltonian in the perturbative high-frequency expansion
\begin{equation}
    H_\mathrm{eff} = \sum_{n=0}^{\infty}H_\mathrm{rot}^n
\end{equation}
where the expansion coefficients are inversely proportional to the driving frequency $H_\mathrm{eff}^n\propto1/\nu^n$. By performing this expansion on the Hamiltonian in the rotating frame, we ensure the convergence of the expansion as long as $h\nu\gg U-\ell h \nu,\,t$.

The resulting effective Hamiltonian is of the form
\begin{equation}
    H_\mathrm{eff} = 
    \begin{pmatrix}
        U_\mathrm{eff}& (-1)^{\ell+1}\cdot t_\mathrm{eff} & (-1)^\ell\cdot t_\mathrm{eff}& V_\mathrm{eff}^\mathrm{ct}\\
        (-1)^{\ell+1}\cdot t_\mathrm{eff}& V_\mathrm{eff}^\mathrm{nn}
        & -V_\mathrm{eff}^\mathrm{de}&-t_\mathrm{eff}\\
     (-1)^\ell \cdot t_\mathrm{eff}&-V_\mathrm{eff}^\mathrm{de}             & V_\mathrm{eff}^\mathrm{nn}              &t_\mathrm{eff}\\
    V_\mathrm{eff}^\mathrm{ct}   &-t_\mathrm{eff} &t_\mathrm{eff}& U_\mathrm{eff}
    \end{pmatrix}
\end{equation}
where the effective Hubbard parameters themselves are expansions of the inverse driving frequency and Bessel functions.
Interestingly, the drive changes the coupling between the eigenstates by the harmonic order dependent sign change to the tunnel coupling.
For odd harmonic order, the singlet couples to $\ket{d_-}$, whereas for even harmonic order $\ket{s}$ couples to $\ket{d_+}$, like in the static scenario. 

\subsection{Effective Hubbard parameters in driven double wells}
In this section we explore the effective Hubbard parameters in the high frequency expansion.
For this purpose, we have considered terms up to the third order in the inverse frequency  $1/\nu$ while incorporating the Bessel functions $\mathcal{J}_{j}(K_0)$ for $j=0,1,2,3$.

The effective Hamiltonian is dominantly characterized by the effective tunneling and effective interaction

\begin{equation}
    \begin{aligned}
        t_\mathrm{eff}^{(\ell)} &=  t\,\mathcal{J}_\ell(K_0) +\mathcal{O}(\nu^{-2})\\
        U_\mathrm{eff} &= U-
    \ell h\nu
    +\frac{t^2}{h\nu}\;\sum_j\alpha_j^{(\ell)}\mathcal{J}_j(K_0)^2\\
    &+\frac{t^2(U-\ell h \nu)}{(h\nu)^2} \;\sum_j\tilde{\alpha}_j^{(\ell)}\mathcal{J}_j(K_0)^2
    +\mathcal{O}(\nu^{-3})
    \end{aligned}
\end{equation}
with the harmonic order dependent pre-factors of the $j$th Bessel function $\alpha_j^{(\ell)},\,\tilde{\alpha}_j^{(\ell)}$ given in table \ref{sup:tab1}.
We compare the extracted effective Hubbard parameters at the trap center (refer to manuscript for details) to our theoretical expectations in figure \ref{fig:fig1}, for an attractively interacting system $U/t\approx-9$, with a driving amplitude $K_0=2.4$.
The effective tunneling $t_\mathrm{eff}^{(\ell)}$ differs between the harmonic orders ($h\nu\sim U$ and $h\nu\sim U/2$) due to the different Bessel function dependence, while there is no clear driving frequency dependency visible.
On the other hand, the effective interaction changes with the driving frequency linearly with different slopes depending on the harmonic order, as theoretically expected.

\begin{figure}
        \centering
        \includegraphics[width=\columnwidth]{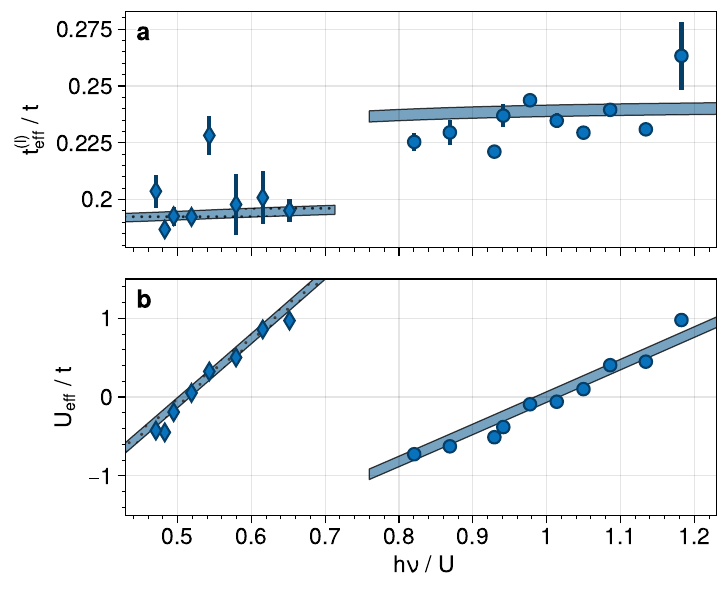}
        \caption{The effective Hubbard parameters are plotted versus the driving frequency $h\nu/U$ for attractive interactions $U/t\approx-9$ and a driving amplitude of $K_0=2.4$. (a) The effective tunneling $t_\mathrm{eff}^{(\ell)}$ differs between the first harmonic order (circles) and the second harmonic order (diamonds) due to the Bessel function dependence $t_\mathrm{eff}^{(\ell)}\propto J_\ell(K_0)$.
        (b) The effective interaction strength $U_\mathrm{eff}$ changes linearly with the driving frequency with a harmonic order dependent slope $U_\mathrm{eff}\approx U -\ell h \nu$.
        }
        \label{fig:fig1}
\end{figure}

The periodic modulation also modifies the higher order Hubbard parameters 
\begin{equation}
    \begin{aligned}
   V_\mathrm{eff}^\mathrm{ct} &=
   \frac{t^2}{h\nu}\;\sum_j\beta_j^{(\ell)}\mathcal{J}_j(K_0)^2\\ 
    &+\frac{t^2(U-\ell h \nu)}{(h\nu)^2} \;\sum_j\tilde{\beta}_j^{(\ell)}\mathcal{J}_j(K_0)^2
    +\mathcal{O}(\nu^{-3})\\
     V_\mathrm{eff}^\mathrm{nn} &=
    \frac{t^2}{h\nu}\;\sum_j\gamma_j^{(\ell)}\mathcal{J}_j(K_0)^2\\
    &+\frac{t^2(U-\ell h \nu)}{(h\nu)^2} \;\sum_j\tilde{\gamma}_j^{(\ell)}\mathcal{J}_j(K_0)^2
    +\mathcal{O}(\nu^{-3})\\
    V_\mathrm{eff}^\mathrm{de} &=  V_\mathrm{eff}^\mathrm{nn}
    \end{aligned}
\end{equation}

with the harmonic order dependent pre-factors of the $j$th Bessel function $\beta_j^{(\ell)},\,\tilde{\beta}_j^{(\ell)}$ and $\gamma_j^{(\ell)},\,\tilde{\gamma}_j^{(\ell)}$ given in table \ref{sup:tab1}.  
We compare these higher order effective Hubbard parameters with the effective tunneling and interaction in figure \ref{fig:fig2}, for two interaction strengths $U/t\approx-9$, $\ell=2$ and $U/t\approx-18$, $\ell=1$. For the latter, the absolute driving frequencies are large which renders the higher-order parameters almost negligible.
In contrast, for the smaller driving frequencies of the smaller interaction strength, the higher order parameters are of considerable amplitude. In particular, for strong repulsive interactions $U_\mathrm{eff}/t_\mathrm{eff}^{(\ell)}\approx6$, the effective correlated tunneling is comparable to the effective tunneling strength $V^\mathrm{ct}_\mathrm{eff}/t_\mathrm{eff}^{(\ell)}\approx\SI{4e-1}{}$.
This shows an enhancement of two orders of magnitude of the relative correlated tunneling compared to the static counterpart.
\begin{figure}
        \centering
        \includegraphics[width=\columnwidth]{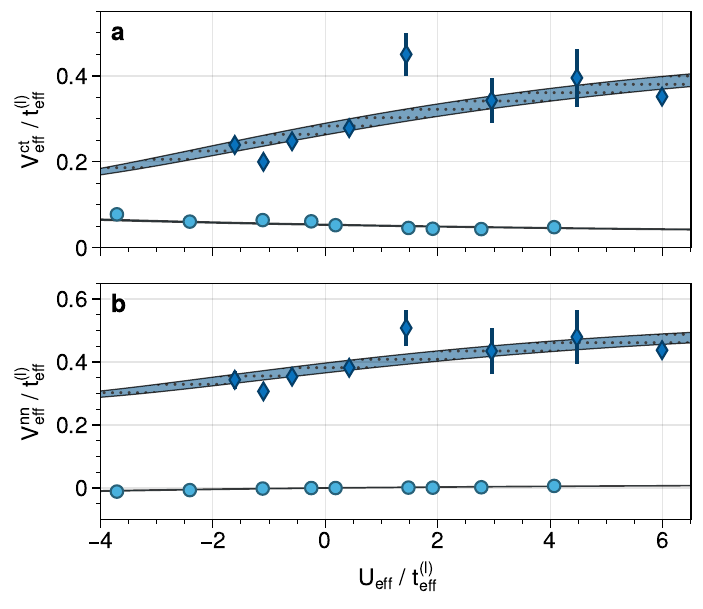}
        \caption{The effective correlated tunneling $V_\mathrm{eff}^\mathrm{ct}$ (a) and nearest-neighbor interaction $V_\mathrm{eff}^\mathrm{nn}$ (b) are plotted versus the effective interaction $U_\mathrm{eff}/t_\mathrm{eff}^{(l)}$ for a driving amplitude $K_0=2.4$ and two interaction strengths $U/t\approx-9$, $\ell=2$ /(blue) and
        $U/t\approx-18$, $\ell=1$ (light blue).
        Note that the direct exchange interaction is always of the same magnitude as the nearest-neighbor interaction $V_\mathrm{eff}^\mathrm{de}=V_\mathrm{eff}^\mathrm{nn}$.
        The data is extracted from the fitted Hamiltonian at the trap center and shown as circles (diamonds) for the first (second) harmonic near-resonant theory.
        }
        \label{fig:fig2}
\end{figure}

Of particular interest is the effective pair tunneling amplitude in the driven system. It is given by the energy difference of the two states with dominantly double-occupancies.
These states change with the harmonic order due to the aforementioned change of coupling which gives the effective pair tunneling amplitude 
\begin{equation}
    J_\mathrm{eff} = \begin{cases}E_{g,e}-E_{d_+} &  \ell\;\text{odd}\\
    E_{g,e}-E_{d_-} & \ell\;\text{even} \quad.
    \end{cases}
\end{equation}
However, the higher-order Hubbard parameters can have significant amplitude in the driven scenario and must therefore be considered for the effective pair tunneling amplitude
\begin{equation}
 \begin{aligned}
        J_\mathrm{eff}
        &= \frac{-U_\mathrm{eff}+(-1)^\ell3V_\mathrm{eff}^\mathrm{ct}+2V_\mathrm{eff}^\mathrm{nn}}{2}\\&\pm
        \frac{\sqrt{16t_\mathrm{eff}^2+U_\mathrm{eff}^2+(-1)^\ell2U_\mathrm{eff}V_\mathrm{eff}
        ^\mathrm{ct}-4U_\mathrm{eff}V_\mathrm{eff}^\mathrm{nn}}}{2}\\
        & \frac{\overline{+
        ((-1)^{\ell+1} V_\mathrm{eff}^\mathrm{ct}+2V_\mathrm{eff}^\mathrm{nn})^2}}{2}
 \end{aligned}
 .
\end{equation}
Interestingly, the effective pair tunneling does not necessarily approach the effective pair superexchange for strong effective interactions $|U_\mathrm{eff}|\gg t_\mathrm{eff},\,V_\mathrm{eff}^\mathrm{ct},\,V_\mathrm{eff}^\mathrm{nn}$:
\begin{equation}
     J_\mathrm{eff} \approx \frac{4t_\mathrm{eff}^2}{U_\mathrm{eff}}+(-1)^{\ell}2V_\mathrm{eff}^\mathrm{ct}+\mathcal{O}(U_\mathrm{eff}^{-2})
\end{equation}
Instead, for effective repulsive interactions the effective pair tunneling is enhanced (suppressed) by the correlated tunneling for even (odd) harmonic order and vice versa for attractive interactions.

\begin{table}
\renewcommand{\arraystretch}{1.2}
\vspace{0.1 in}
\begin{center}
\begin{tabular}{|c|cc|cc|}
\hline
Pre-Factor&\multicolumn{2}{c|}{l=1}&\multicolumn{2}{c|}{l=2}\\
&$j=0$&$j=1$&$j=0$&$j=1$\\

\hline

$\alpha_j^{\ell}$& 2 & 1 &1 &8/3 \\

$\tilde{\alpha}_j^{\ell}$& -2 & -1/2 &-1/2 &-20/9\\
\hline
$\beta_j^{\ell}$& 2 & -1 &1 &-8/3\\
$\tilde{\beta}_j^{\ell}$& -2 & -1/2 &-1/2 &20/9\\
\hline
$\gamma_j^{\ell}$& -2 & -1 &-1&-8/3\\
$\tilde{\gamma}_j^{\ell}$& 2 & 1/2 &1/2&20/9\\
\hline
\end{tabular}
\end{center}
\caption{Bessel function pre-factors of high-frequency expansion of effective Hubbard parameters}
\label{sup:tab1}
\end{table}

\end{document}